\documentclass[11pt]{article}
\usepackage{amsmath}
\usepackage{amsfonts}
\usepackage{amssymb}
\usepackage{graphicx}
\begin{document}
\title{\bf The driven oscillator, with friction}
\author{T.B. Smith
\\
\ {\scriptsize School of Physical Sciences}\\{\scriptsize The Open University, Walton Hall, Milton Keynes, MK7 6AA, UK}}
\maketitle
\begin{abstract}
This paper develops further the semi-classical theory of an harmonic oscillator acted on by a Gaussian white noise force discussed in \cite{smith15} (arXiv:1508.02379 [quant-ph]).  Here I add to that theory the effects of Brownian damping (friction).  This requires an adaption of the original formalism and complicates the algebra somewhat. Albeit semi-classical, the theory can be used to model quantum expectations and probabilities. Among several examples, I consider some implications for the canonical phase operator.
\end{abstract}


\section{Introduction}
In the classical theory of Brownian motion in phase space a particle is subjected both to a white noise external force and to a damping force proportional to its velocity \cite{smith79,wax}.  To generate a semiclassical theory from this for an oscillator we make the association $(p,q)\rightarrow (\hat{p},\hat{q})$ where $[\hat{p},\hat{q}]=-{\rm i}\hbar$, but the damping term makes the transition to quantum dynamics awkward.  The solution has been known for some time \cite{kerner,havas1,havas2}---there is an appropriate time evolution Hamiltonian which is not the energy. This is the topic of section \ref{WignerWeyl} wherein the classical dynamics is transcribed to the Wigner-Weyl quantization formalism.

Section \ref{noiseforcing} adds ensemble statistics to the Wigner-Weyl time propagator, thereby incorporating Brownian statistics semi-classically.  An exact expression for the propagator of the density matrix in the Wigner-Weyl picture is given.  At long times it forgets its history and becomes thermal.

In section \ref{transprob} we consider the transition probability of the oscillator from the harmonic oscillator ground state. The Wigner-Weyl formalism is used throughout. An exact expression for this probability was worked out in \cite{smith15} when there is strictly no friction in which case the oscillator cannot thermalise. In the present case, with non-zero friction, the system {\em can} thermalise, but the multiple integrals involved in many cases, though often Gaussian in form, are somewhat lengthy so I resort to computation.  The plot of one example is shown in Figure 1.

Section \ref{phase} discusses the `canonical phase operator', $\hat{\phi}$, which is, roughly speaking, the Weyl quantization of $\arctan (p/q)$ where $p$ and $q$ are the canonical coordinates in appropriate units.  Also briefly discussed is what might be termed the `physical' phase operator, $\hat{\overline{\phi}}$, which is, again roughly speaking, the Weyl quantization of  $\arctan (m \dot{q}/q)$.  Should the oscillator initially be in the ground state, Figure 2 shows, by computation, how the expectation of the canonical phase operator decays to zero as time advances. Section \ref{phase} also considers the spectra of both phase operators, again by computation, Figure 3. To my knowledge, explicit expressions for the spectral representations of these operators haven't been given.  Finally, Section \ref{phase} also considers the variance of both angle operators in the thermal limit.

Section \ref{R} briefly considers in the thermal limit the expectation \[\overline{{\rm Tr}\big( \hat{\rho}(t){\rm exp}(-B \hat{E}_{\rm osc})  \big)}\] where $\hat{E}_{\rm osc}$ is the operator for the physical energy of the oscillator and $\hat{\rho}(t)$ is an arbitrary state.  In the long time limit the result is, perhaps unsurprisingly, classically thermal.

Section \ref{disc} gives a brief discussion.
\section{State evolution in the Wigner-Weyl picture}\label{WignerWeyl}
There are many possible formulations of quantum mechanics in phase space \cite{cohen}.
Generally, they can be related \cite{smith06,lee} to that of Wigner and Weyl \cite{weyla,weylb,wigner}.
As in \cite{smith15}, where details are given, we shall adopt a formal efficient notation \cite{degroot}.
Denoting the Weyl transform of an operator $\hat{A}$ by $A(p,q)$, or sometimes by $(\hat{A})(p,q)$, it is given by
\begin{equation}\label{weyl1}
  \hat{A} \longleftrightarrow {\rm Tr}(\hat{A}\,\hat{\Delta}(p,q))\equiv A(p,q)\equiv (\hat{A})(p,q)
\end{equation}
where
\begin{eqnarray}\label{weyl}
\mathrm{}    \hat{\Delta}(p,q) & = &
    \int_{-\infty}^{\infty}\frac{ {\rm d} p' {\rm d}q'}{h}
    \,{\rm e}^{-\frac{\rm i}{\hbar} (p'q - q'p )}
    \,\hat{D}(p',q') \nonumber\\
    \\
    & = & \int_{-\infty}^{\infty}
    {\rm d}x \,\,{\rm e}^{\frac{\rm i}{\hbar}p\, x}\,\,
   |q + \frac{x}{2} \rangle \langle q - \frac{x}{2}|  \nonumber
\end{eqnarray}
and $\hat{D}$ is the Weyl operator (\cite{klauder}),
\begin{equation}\label{Dpq}
    \hat{D}(p,q) = {\rm e}^{\frac{\rm i}{\hbar}(p\, \hat{q} - q
\,\hat{p})}\,.
\end{equation}
In this formalism, it is important to realize that $\hat{p}$ and $\hat{q}$ are canonical operators such that $[\hat{p},\hat{q}]=-{\rm i}\hbar$.
Formally $\hat{\triangle}$ has the the properties \cite{degroot} that its trace
is unity, that
\begin{equation}\label{weyl3}
    \int_{-\infty}^{\infty}\frac{{\rm d}p\,{\rm d}q}{h}
    \hat{\Delta}(p,q) = 1\,,
\end{equation}
and that
\begin{equation}\label{weyl4}
    {\rm Tr} \left( \hat{\Delta}(p,q) \hat{\Delta}(p',q') \right)
    = h \delta(p - p') \delta(q - q')\,.
\end{equation}
Then equation (\ref{weyl1}) can be inverted to give
\begin{equation}\label{weylone}
    \hat{A} = \int_{-\infty}^\infty \frac{{\rm d}p\,{\rm d}q}{h}
    A(p,q) \hat{\Delta}(p,q) \quad\mbox{and}\quad {\rm Tr}\hat{A}=\int_{-\infty}^\infty \frac{{\rm d}p\,{\rm d}q}{h}
    A(p,q)\,.
\end{equation}
From these properties one can also show \cite{degroot} that, for two operators
$\hat{A}$ and $\hat{B}$,
\begin{equation}\label{weyl5}
    {\rm Tr}(\hat{A}\hat{B}) = \int_{-\infty}^{\infty}\frac{{\rm d}p\, {\rm d}q}{h}
    A(p,q) B(p,q)\,,
\end{equation}
and that the Weyl transform of the product $\hat{A} \hat{B}$ is
\begin{equation}\label{weyl6}
 \hat{A} \hat{B}  \longleftrightarrow {\rm Tr}\left(\hat{A} \hat{B} \hat{\Delta}(p,q) \right)=
    A(p,q)\, {\rm exp}\left[ \frac{{\rm i}\hbar}{2}
    \left( \frac{\partial^*}{\partial q} \frac{\partial}{\partial p}\, -\,
     \frac{\partial^*}{\partial p} \frac{\partial}{\partial q} \right) \right] B(p,q)\,,
\end{equation}
where the starred operators act to the left on $A(p,q)$.

The Weyl transform makes the sensible fundamental associations
$\hat{1} { \longleftrightarrow} 1$, $F(\hat{p}){ \longleftrightarrow} F(p)$
\\and $F(\hat{q}) {\longleftrightarrow} F(q)$,
but functions that mix $\hat{p}$ and $\hat{q}$ are more complicated. For instance\\ $\hat{p}\, \hat{q}{ \longleftrightarrow} p\,q -{\rm i}\hbar/2 $ and $\hat{q} \hat{p}{ \longleftrightarrow} p\,q +{\rm i}\hbar/2 $ so that the Weyl operator corresponding to $p \, q$ \\is $(\hat{p}\, \hat{q}+ \hat{q}\, \hat{p})/2 $.

The classical equation of motion for a damped harmonic oscillator forced by $F(t)$ is
\begin{equation}\label{eqnq}
  m\ddot{q}+m \beta \dot{q}+m \omega^2 q=F(t).
\end{equation}
The time generator, but not the energy, of this system is \cite{kerner,havas1,havas2} the Hamiltonian
\begin{equation}\label{timegen}
  H_t(p,q)=\frac{p^2}{2m}\,{\rm e}^{-\beta t}+{\rm e}^{\beta t}\left(\frac{1}{2}m \omega^2 q^2 - q F(t)\right),
\end{equation}
To verify that this is the correct Hamiltonian we note that
\[ \dot{q}=\frac{\partial H_t}{\partial p}=\frac{p}{m}{\rm e}^{-\beta t}\quad \mbox{and}\quad
\dot{p}=-\frac{\partial H_t}{\partial q}={\rm e}^{\beta t}(F(t)-m \omega^2 q), \]
from which it follows that \[m \ddot{q}={\rm e}^{- \beta t}(\dot{p}-\beta p), \]
so leading to equation (\ref{eqnq}).

The Lagrangian corresponding to $H_t(p,q)$ is
\[L(q,\dot{q})=p \dot{q}-H_t(p,q) \]
so that the canonical variables $(p,q)$ in phase space are
\begin{equation}\label{phasevars}
p=\frac{\partial L}{\partial \dot{q}}=m\dot{q}\,{\rm e}^{\beta t}\equiv P {\rm e}^{\beta t} \,\,{\rm and}\,\,q,
\end{equation}
where we use the symbol $P$ to denote $m \dot{q}$, the physical momentum.
For the Wigner-Weyl association the operator governing the quantum time dependence of this system is
\begin{equation}\label{Ht}
 \hat{H}_t=\frac{\hat{p}^2}{2m} {\rm e}^{- \beta t}+
{\rm e}^{\beta t}\left( \frac{1}{2}m \omega^2 \hat{q}^2 -\hat{q} F(t) \right).
\end{equation}
The the Wigner function is defined \cite{degroot} as
\begin{equation}\label{wigt}
  \rho_w(p,q;t)\equiv \frac{1}{h} {\rm Tr}\big(\hat{\rho}(t) \hat{\Delta}(p,q) \big)
\longleftrightarrow \frac{1}{h}\hat{\rho}(t)\,,
\end{equation}
so that by (\ref{weyl5}),
\begin{equation}\label{wignert}
  {\rm Tr} \big(\hat{\rho}(t) \hat{A}\big) = \int_{-\infty}^{\infty} {\rm d}p \, {\rm d}q \, \rho_w(p,q;t) A(p,q)\,.
\end{equation}
The evolution of the Wigner function under the action of a time-dependent Hamiltonian is is well-known \cite{smith06,degroot,smith78}. Wave functions evolve according to
\begin{equation}\label{wftime}
  |\psi_t \rangle = \hat{U}_t | \psi\rangle,
\end{equation}
where the unitary time evolution operator ${\hat U}_t$ is governed by the equation
\begin{equation}\label{evo}
  {\rm i} \hbar \frac{\partial}{\partial t} \hat{U}_t =
  \hat{H}_t \hat{U}_t\,,
\end{equation}
where, in this case, the Hamiltonian is $\hat{H}_t$, equation (\ref{Ht}).
Then the time-dependent density matrix $\hat{\rho}(t)$ is given by
\begin{equation}\label{opev}
 \hat{\rho}(t) = {\hat U}_t \, \hat{\rho}(0)\,{\hat U}^{\dagger}_t\,,
\end{equation}
and its Weyl transform is
\begin{equation}\label{opevo}
    \rho_w(p,q;t) = \int {\rm d}p' {\rm d}q' P_w(p,q,t|p',q',0)\rho_w(p',q';0)
\end{equation}
where $P_w(\cdot|\cdot)$ is the Wigner propagator defined by
\begin{equation}\label{wignerprop}
    P_w(p,q,t|p',q',0) = \frac{1}{h}{\rm Tr}(\hat{U}_t^\dag \hat{\Delta}(p,q)\hat{U}_t \hat{\Delta}(p',q')).
\end{equation}
In particular, it is easy to show from the properties above that
\begin{equation}\label{property1}
    \int{\rm d}p' {\rm d}q' P_w(p,q,t|p',q',0)=\int{\rm d}p\, {\rm d}q P_w(p,q,t|p',q',0)=1
\end{equation}
and
\begin{equation}\label{property2}
    P_w(p,q,0|p',q',0) = \delta(p-p')\delta(q-q')\,.
\end{equation}
From definition (\ref{wignerprop}) and equation (\ref{evo}) we can differentiate the propagator with respect to time to get
\begin{equation}\label{commutator}
  {\rm i}\hbar \frac{\partial}{\partial t}P_w(p',q',t|p,q,0)=
\frac{1}{h}
 \left( \, [\hat{H}_t,{\hat U}_t \hat{\Delta}(p,q) {\hat U}_t^\dag]\, \right )(p',q')
\end{equation}
The Weyl transform of ${\hat H}_t$, equation (\ref{timegen}), is
quadratic in $p$ and $q$ and so has no derivatives with respect to $p$ and/or $q$ of order higher than second.  Applying equation (\ref{weyl6}) to equation (\ref{commutator}) is straightforward. Collecting terms gives
\begin{eqnarray}\label{propmotion}
    \lefteqn{\frac{\partial}{\partial t}P_w(p,q,t|p',q',0) = }\nonumber\\ & = & \frac{\partial}{\partial q}H_t(p,q)\frac{\partial}{\partial p}P_w(p,q,t|p',q',0)-\frac{\partial}{\partial p}H_t(p,q)
    \frac{\partial}{\partial q}P_w(p,q,t|p',q',0)\,.
\end{eqnarray}
This describes classical motion, under the action of $H_t(p,q)$, of the canonical variables $p$ and $q$, equation (\ref{phasevars}). The solution that satisfies initial condition (\ref{property2}) is
\begin{equation}\label{classicalP}
  P_w(p,q,t|p',q',0) = \delta(p - p(t|p',q',0))\delta(q-q(t|p',q',0))\,,
\end{equation}
where $\big(p(t|p',q',0),q(t|p',q',0)\big)$ is the classical phase space solution for canonical momentum and position under the action of Hamiltonian $H_t(p,q)$ such that $\big(p(0|p',q',0),q(0|p',q',0)\big)= (p',q')$.  This solution also obeys equation (\ref{property1}), by direct integration in the first instance and, in the second, by recognizing that the Jacobian $\partial(p,q)/ \partial(p',q')$ is unity when $(p,q)$ and $(p',q')$ are related by the classical motion implied by equation (\ref{classicalP}).  For the cases we are considering, the equations for classical motion are
\begin{equation}\label{motion}
  p(t) = m\dot{q}(t)\, {\rm e}^{\beta t} \quad \mbox{and}\quad m\ddot{q}(t)+m \beta \dot{q}(t)+m \omega^2 q(t)=F(t).
\end{equation}
\section{Forcing by stationary white noise, with friction added}\label{noiseforcing}
The white noise force is a stationary Gaussian process \cite{wax} with the particular ensemble averages,
\begin{equation}\label{noise}
  \overline{F(t)} = 0 \quad\mbox{and}\quad \overline{ F(t_1)F(t_2)} = \mu\, \delta(t_1 - t_2)\,.
\end{equation}
With friction added this describes the classical theory of Brownian motion \cite{wax}.
The friction term is $m \beta \dot{q}$ in (\ref{motion}) and represents phenologically the average effect of a heat bath. In our semi-classical theory the effects of Brownian motion are expressed by taking the ensemble average of the propagator (\ref{classicalP}).  Denoting this average by an overline, from equation (\ref{classicalP}) we have
\[\overline{ P_w{(p,q,t|p_0,q_0,0)}}=\overline { \delta(p-p(t|p_0,q_0,0)) \delta(q-q(t|p_0,q_0,0))} \equiv W(p,q,t|p_0,q_0,0)  \]
where $W(\cdot|\cdot)$ is the conditional probability density for $(p,q)$ at time $t$ given the initial conditions $(p_0,q_0)$ at $t=0$, and one must remember that $p$ and $q$ are the
canonical variables (\ref{phasevars}). The connection with the physical variables is straightforward, for
from (\ref{phasevars}) we can write
\begin{eqnarray}\label{newprop}
  W(p,q,t|p_0,q_0,0) &=& \overline{\delta(p-m \dot{q}(t)\,{\rm e}^{\beta t}) \delta(q-q(t)}\nonumber \\
  \\
   &=& {\rm e}^{-\beta t} \,\overline{\delta(p\, {\rm e}^{-\beta t}-m \dot{q}(t))\delta(q-q(t))}\nonumber,
\end{eqnarray}
with initial conditions $q(0)=q_0$ and $m\dot{q}(0)=p_0$.
Classical Brownian motion is random, stationary and Markovian, and can be characterized by the conditional probability density $W(P,q,t|P_0,q_0,0)$ where $P=m \dot{q}$ is the physical momentum. Reference \cite{smith79} gives the following efficient expression that is generally approximate, but is \emph{exact} when the energy is at most quadratic in $P$ and $q$:
\begin{eqnarray}\label{brownianone}
    \lefteqn{W(P,q,t|P_0,q_0,0) \simeq  \int \frac{{\rm d}a{\rm d}b}{(2 \pi)^2}\,
  {\rm exp}\big({\rm i}a(q-q(t)\big){\rm exp}\big({\rm i}b(P-P(t)\big)}\nonumber\\
   && \hspace{5cm} \times \,\, {\rm exp}\left[-\frac{\mu}{2}  \int_0^t {\rm d}s
  \Big(\partial_{P_s}\big(aq(t)+bP(t)\big)\Big)^2\right]\hspace{4cm}
\end{eqnarray}
where $\mu$ characterizes the random force $F(t)$, equation (\ref{noise}). In (\ref{brownianone}) $(P(t),q(t))$ is shorthand for solutions $\big(P(t| P_0,q_0,0),q(t| P_0,q_0,0)\big)$ (where $P(t)=m\dot{q}$), $\dot{q}$ follows from the solution to (\ref{motion}), and $\partial_{P_s}$ is the partial derivative with respect to the physical momentum $P_s$ at the intermediate time $0 \leq s\leq t$.  Thus $\mu$ occurs explicitly in equation (\ref{brownianone}) and $\beta$ occurs implicitly through the solution $(P(t),q(t))$. If the classical Brownian particle were to thermalize after long times \cite{smith79} then\footnote{This corrects a typographical error in Section 4.2 of \cite{smith15}.} the relation  $\mu/2=m\beta \Theta $ must obtain, where $\Theta=k T$, with Boltzmann's constant $k$ and temperature $T$.
In terms of canonical variables $p$ and $q$ the required propagator is
\begin{equation}\label{canprop}
  W(p,q,t|p_0,q_0,0)={\rm e}^{- \beta t}W({\rm e}^{- \beta t}p,q,t|p_0,q_0,0)
\end{equation}

In our picture we require a solution, where $\beta$ is nonzero, for the classical unforced harmonic oscillator, $P(t)=m \dot{q}(t)=P_t=P(t|P_s,q_s,s)$ and $q(t)=q_t=q(t|P_s,q_s,s)$. For the underdamped case ($\omega \geq \beta/2 $) these are, for $0\leq s \leq t$,
\begin{equation}\label{clone}
  P_t={\rm e}^{-\frac{\beta (t-s)}{2}}\big[P_s \big(\cos\Omega(t-s)-\frac{\beta}{2 \Omega}\sin\Omega (t-s)\big)-q_s \frac{m \omega^2}{\Omega} \sin\Omega(t-s)\big]
\end{equation}
and
\begin{equation}\label{cltwo}
  q_t = {\rm e}^{-\frac{\beta (t-s)}{2}}\big[q_s \big(\cos\Omega(t-s)+\frac{\beta}{2 \Omega}\sin\Omega (t-s)\big)+P_s \frac{1}{m \Omega} \sin\Omega(t-s)\big],
\end{equation}
where
\begin{equation}\label{Omega}
  \Omega^2 = \omega^2 - \frac{\beta^2}{4}\,.
\end{equation}
At this point it is convenient to transform from canonical phase space coordinates $(p,q)$ to the dimensionless coordinates $(x,y)$, such that
\begin{equation}\label{coordns}
  (x,y)\equiv (\frac{p}{\hbar \alpha},\alpha q)\hspace{.2cm}{\rm and}\hspace{.2cm}(X,y)\equiv (\frac{P}{\hbar \alpha},\alpha q)=({\rm e}^{-\beta t}x,y),\hspace{.2cm} {\rm where} \hspace{.2cm} \alpha^2=\frac{m \omega}{\hbar}.
\end{equation}
Thus we can rewrite equations (\ref{clone}) and (\ref{cltwo}) as
\begin{equation}\label{CLone}
  X_t=\frac{P_t}{\hbar \alpha}={\rm e}^{-\frac{\beta (t-s)}{2}}\big[X_s \big(\cos\Omega(t-s)-\frac{\beta}{2 \Omega}\sin\Omega (t-s)\big)-y_s \frac{\omega}{\Omega} \sin\Omega(t-s)\big],
\end{equation}
and
\begin{equation}\label{CLtwo}
  y_t = {\rm e}^{-\frac{\beta (t-s)}{2}}\big[y_s \big(\cos\Omega(t-s)+\frac{\beta}{2 \Omega}\sin\Omega (t-s)\big)+X_s \frac{\omega}{\Omega} \sin\Omega(t-s)\big],
\end{equation}
Using this information to evaluate (\ref{brownianone}) and replacing $\simeq$ by $=$, for this result is exact \cite{smith79}, gives
\begin{eqnarray}\label{osc}
  \lefteqn{W(x,y,t|x_0,y_0,0) ={\rm e}^{- \beta t}  \int \frac{{\rm d}a{\rm d}b}{(2 \pi)^2 \hbar}\,
  {\rm exp}\big[{\rm i}a \epsilon(y-y_t)\big]}\nonumber\\
   &&\hspace{1cm}\times \,{\rm exp}\big[ {\rm i}\frac{b}{\epsilon}(x {\rm e}^{- \beta t}-X_t) \big]\, {\rm exp}\left[-\frac{N}{2}\int_0^{\Omega t} {\rm d}\theta \,{\rm e}^{-\frac{\beta}{\Omega}\theta}\Big(a \sin \theta+ b( \cos \theta -\frac{\beta}{2 \Omega} \sin\theta ) \Big)^2 \right],
\end{eqnarray}
where $(X_t,y_t)$ are given by (\ref{CLone}) and (\ref{CLtwo}) with initial time $s=0$, $(x_0,y_0)= (\frac{p_0}{\hbar \, \alpha},\alpha  q_0)$, $N=\mu /(m \Omega^2 \hbar)$, and $\epsilon=\sqrt{\Omega/\omega}$. This expression for $W(|)$ is normalised with respect to integration over ${\rm d}p{\rm d}q =\hbar\,{\rm d}x{\rm d}y$.

The integral in (\ref{osc}) has a two-dimensional Gaussian form and thus can be evaluated to give another Gaussian form in $x$ and $y$. In particular, consider the long-time limit, for which $\beta t $ is not small. Ignoring terms damped by the factor ${\rm exp}(- \beta t)$,
\begin{equation}\label{longtime}
 \frac{N}{2}\int_0^{\Omega t} {\rm d}\theta \,{\rm e}^{-\frac{\beta}{\Omega}\theta}\Big(a \sin \theta+ b( \cos \theta -\frac{\beta}{2 \Omega} \sin\theta ) \Big)^2 \approx N \frac{\Omega^3}{4 \omega^2 \beta}(a^2 + b^2 \frac{\omega^2}{\Omega^2})\,.
\end{equation}
Using this, and ignoring the terms $y_t$ and $X_t$ in (\ref{osc}) in this long time limit, gives the product of Gaussian integrals separately with respect to $a$ and $b$. When converted to canonical variables $p$ and $q$ via (\ref{coordns}) and with the choice
\begin{equation}\label{mu}
  \mu = 2 m \beta \Theta\quad {\rm where}\quad\Theta = k T
\end{equation}
the result is that, for long times,
\begin{equation}\label{thermalxy}
W(x,y,t|x_0,y_0,0)\approx \frac{1}{2 \pi \hbar} \frac{\hbar \omega}{\Theta} {\rm e}^{- \beta t}
  {\rm e}^{-\frac{\hbar \omega}{2 \Theta}\big(x^2 {\rm e}^{-2 \beta t}+y^2\big)}\quad\quad({\rm thermal\,\, limit}),
\end{equation}
where $x$ and $y$ are given by (\ref{coordns}) in terms of canonical coordinates $(p,q)$.  This expression is normalized with respect to integration over $\hbar\,{\rm d}x{\rm d}y$.  Writing it in terms the physical variables $P=m \dot{q} (t)= \hbar \, \alpha x\,{\rm e}^{-\beta t}$ and $q$ shows it to be the Maxwell-Boltzmann distribution for the oscillator.

That expression (\ref{thermalxy}) is Maxwell Boltzmann underlines that this theory is semi-classical. It is a single particle with a c-number term to describe the forces acting. A more fully quantum theory would involve interaction with a heat bath \cite{Hondaetc,PGetc}.  Notwithstanding its semi-classicality, in the following I give examples of how it can be used to model quantum effects.
\section{Transition probabilities}\label{transprob}
The probability for transition between states $|\psi_1\rangle$ and $|\psi_2\rangle$ is
\begin{eqnarray}\label{trans}
  \hspace{-1cm}|\langle \psi_2 |\hat{U}_t |\psi_1 \rangle|^2 &=& { \rm Tr}\left(|\psi_2\rangle\langle\psi_2|\hat{U}_t|\psi_1\rangle\langle \psi_1|\hat{U}_t^\dag \right)\nonumber\\
  &=& \int \frac{{\rm d}p\,{\rm d}q}{h}\int {\rm d}p'{\rm d}q'\, \big(|\psi_2\rangle\langle\psi_2|\big)(p,q)P_w (p,q,t|p',q',0)\big(|\psi_1\rangle\langle\psi_1|\big)(p',q')\,,
\end{eqnarray}
where $P_w$ is the Wigner propagator, (\ref{wignerprop}). In particular, for the random driving force (\ref{noise}), the ensemble averaged propagator is
\begin{equation*}
  W (p,q,t|p',q',0)=\overline{P_w(p,q,t|p',q',0)}\,.
\end{equation*}
The corresponding ensemble averaged transition probability is
\begin{equation}\label{average}
 \overline{ |\langle \psi_2 |\hat{U}_t |\psi_1 \rangle|^2} = \int \frac{{\rm d}p\,{\rm d}q}{h}\int {\rm d}p'{\rm d}q'\, \big(|\psi_2\rangle\langle\psi_2|\big)(p,q)W (p,q,t|p',q',0)\big(|\psi_1\rangle\langle\psi_1|\big)(p',q')
\end{equation}
where, for an oscillator, $W$ is given by (\ref{osc}). Translated to variables ($x,y$) this is
\begin{equation}\label{averagexy}
 \overline{ |\langle \psi_2 |\hat{U}_t |\psi_1 \rangle|^2} = \int \frac{{\rm d}x\,{\rm d}y}{2 \pi}\int \hbar \,{\rm d}x'{\rm d}y'\, \big(|\psi_2\rangle\langle\psi_2|\big)(x,y)W (x,y,t|x',y',0)\big(|\psi_1\rangle\langle\psi_1|\big)(x',y')
\end{equation}
As in \cite{smith15} we might suppose the oscillator were initially in the ground state, so that $|h_0\rangle$, where
\begin{equation}\label{groundstate}
  \langle \xi|h_0\rangle= (\alpha/ \sqrt{\pi})^{1/2}{\rm exp}\left(- \alpha^2 \xi^2/2\right).
\end{equation}
and ask for the probability $\overline{|\langle h_0 |\hat{U}_t |h_0 \rangle|^2 }$ that it stays there.
Now
\begin{equation}\label{ground}
  \left(|h_0 \rangle\langle h_0|\right)(x,y)=2 \,{\rm exp}\left(- (x^2 +y^2)\right)=2\, {\rm exp}(-R^2)\,
\end{equation}
so that from this and (\ref{osc}) it is clear that the evaluation of this probability requires a number of Gaussian integrals only.  This is easy in the limit of long times.  Using (\ref{thermalxy}) and (\ref{ground}) in (\ref{averagexy}) gives, for large $\beta t$,                                                                                                                          %
\begin{equation}\label{problongt}
  \overline{|\langle h_0 |\hat{U}_t |h_0 \rangle|^2 } \approx  {\rm e}^{- \beta t}\,\,\frac{\hbar \omega}{\Theta}
  \sqrt{\frac{1}{(1+\frac{\hbar \omega}{2 \Theta})(1+\frac{\hbar \omega}{2 \Theta}{\rm e}^{- 2\beta t})}}.
\end{equation}
This equation applies for long times it will not, of course, be expected to equal unity at $t=0$.

By contrast, when $\beta$ strictly vanishes thermalization does not occur. For that case we found \cite{smith15} that for all times,
\begin{eqnarray}\label{quadfour}
  \overline{|\langle h_0 |\hat{U}_t |h_0 \rangle|^2 } = \frac{1}{\sqrt{\left(1 + \frac{N_o \,\omega t}{2}\right)^2 -\left(\frac{N_o}{2}\right)^2 \sin^2 \omega t}}\,,
\end{eqnarray}
where $N_o=\mu_o /(m \omega^2 \hbar)$ and $\mu_o$ is a parameter characteristic of the white noise forcing strength and is not necessarily equal to $2 m \beta k T $.

\begin{figure}[h]
\begin{center}
\includegraphics{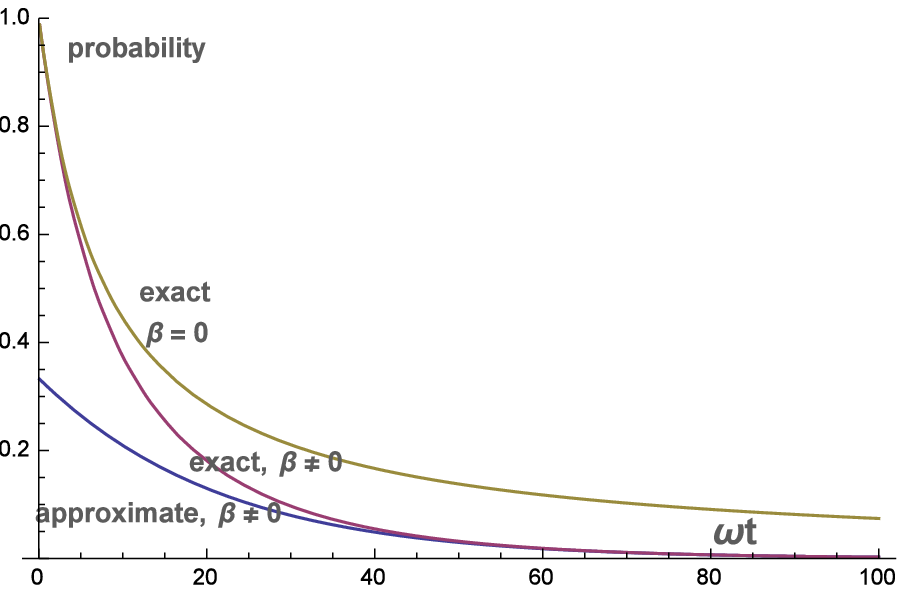}
\caption{\label{avephi}  The probability that the oscillator stays in the ground state, calculated by a series of numerical Gaussian integrals. The lower curve uses the long time ($\beta t\gg 0$) approximation (\ref{thermalxy}) for the propagator. The middle curve is exact with parameter values $D=\frac{2 \theta}{\hbar\omega}=5$ and $B = \frac{\beta}{\omega}=0.05$.  For the top curve, $\beta$ strictly vanishes, equation (\ref{quadfour}), with $N_o\equiv\frac{\mu}{m \omega^2 \hbar}=0.25.$  This result was derived in \cite{smith15}.}
\end{center}
\end{figure}

More generally, when the oscillator is initially in the ground state it is clear from (\ref{osc}), (\ref{ground}) and (\ref{averagexy}) that an exact evaluation of
$\overline{|\langle h_0 |\hat{U}_t |h_0 \rangle|^2 }$  requires the evaluation of several Gaussian integrals. As an example, the result for the parameter values $D=\frac{2 \theta}{\hbar\omega}=5$ and $B = \frac{\beta}{\omega}=0.05$ is shown by the middle curve of Figure \ref{avephi}.
\section{Phase}\label{phase}
\subsection{Definition}\label{phasedef}
In this subsection I consider briefly the generalization of the model in \cite{smith15} to include the effect of friction $\beta$ on the time evolution of the Weyl quantized phase of the oscillator.  In particular, if the creation operator,
 \[ \hat{a}^\dag = \frac{1}{\sqrt{2}}\left(\alpha\hat{q} - {\rm i}\frac{\hat{p}}{\alpha \hbar}\right)\]
has Weyl transform (where $\alpha^2 \equiv m \omega/\hbar.$)
\begin{equation}\label{phasedef1}
  a^*(p,q)=\frac{1}{\sqrt{2}}\left(\alpha q - {\rm i}\frac{p}{\alpha \hbar}\right)= -\frac{{\rm i}}{\sqrt{2}}(x+{\rm i}y)
=\frac{-\rm i}{\sqrt{2}}R\, {\rm e}^{{\rm i} \phi}
\end{equation}
then the canonical phase operator $\hat{\phi}$ can be defined \cite{sdh92} as the Weyl quantization of $\phi$. Properties of $\hat{\phi}$ have been considered previously \cite{dhsbook,sdh92,dhsrims1,lynchrev}. It is a bone-fide bounded self-adjoint operator on Hilbert Space. As befits an angle, its spectrum must be limited to a range of $2 \pi $ which I shall take as $[-\pi,\pi)$.

Generally in the Wigner-Weyl picture, the time-dependent average of an operator $\hat{A}$ with respect to the state $\hat{\rho}$ is given by (\ref{wignert}) with (\ref{opevo}), where, for an oscillator, the evolution is classical, equation (\ref{classicalP}). In particular the Weyl quantizated phase operator is $\hat{\phi}{\longleftrightarrow} \phi(p,q)$, where $\phi(p,q)$ is the harmonic oscillator phase in the plane $(p,q)$. Then, where
${\rm d}p \,{\rm d}q/h = R {\rm d}R {\rm d}\phi / 2 \pi$,
\begin{equation}\label{phi1}
    \phi = {\rm Tr}(\hat{\phi}\,\hat{\Delta}(R,\phi)){ \longleftrightarrow}
    \hat{\phi} = \int_0^\infty{\rm d}R R
\int_{-\pi}^\pi\frac{{\rm d}\phi}{2 \pi}\,\phi\,
\hat{\Delta}(R,\phi)\,,
\end{equation}
where $\phi$ is given in (\ref{phasedef1}) and we may consider $\hat{\Delta}$ as a function of plane polar
coordinates $R$ and $\phi$.  Details are given in \cite{smith15}.

Alternatives to $\hat{\phi}$ have been suggested to represent the phase of an harmonic oscillator, or in some sense photons, for instance by a non-projective positive operator valued measure, or POVM \cite{pegg,pellon1,pellon2,hel,DKPY,KJ}. Using a POVM to represent a quantum system may be thought of as allowing for an element of imperfection in the measurement.
\subsection{Angle operators}\label{phi}
Translating (\ref{wignert}) and (\ref{opevo}) to the variables $(x,y)$, combining them and taking the ensemble average gives, for an operator $\hat{O}$
\begin{equation}\label{aveop}
  \overline{{\rm Tr}(\hat{\rho}(t) \hat{O})}=\int \frac{{\rm d}x\,{\rm d}y}{2 \pi}\int \hbar \,{\rm d}x'{\rm d}y'\, O(x,y)\,W (x,y,t|x',y',0)\,\rho_w(x',y';0),
\end{equation}
where $\rho_w(x',y';0)$ is an arbitrary initial state.  At long times, according to equation (\ref{thermalxy}) memory of the initial state is lost and
\begin{equation}\label{longtime1}
  \overline{{\rm Tr}(\hat{\rho}(t) \hat{O})}\longrightarrow \frac{\hbar \omega}{\Theta} {\rm e}^{- \beta t}
  \int \frac{{\rm d}x\,{\rm d}y}{2 \pi}O(x,y){\rm e}^{-\frac{\hbar \omega}{2 \Theta}\big(x^2 {\rm e}^{-2 \beta t}+y^2\big)}.
\end{equation}
 Writing the Weyl transformation of the operator $\hat{O}$ in polar coordinates as $M(R,\phi)$, say, then,
\begin{equation}\label{longtime2}
  \overline{{\rm Tr}(\hat{\rho}(t) \hat{O})}\longrightarrow \int_{- \pi}^\pi\frac{{\rm d}\phi}{2 \pi}
  \int_0^\infty {\rm d}u \,\, M\Big({\rm e}^{\beta t/2} \sqrt{u \frac{2 \Theta }{\hbar\omega}},\phi\Big)
  {\rm e}^{-u \left( {\rm e}^{- \beta t} \cos^2\phi+{\rm e}^{\beta t}sin^2\phi \right)}\,.
\end{equation}
Now consider those operators, $\hat{\Phi}$, whose Weyl quantizations are functions $\Phi(\phi)$ of phase angle only, namely $\hat{\Phi}{ \longleftrightarrow}\Phi(\phi)$.  For these operators, from (\ref{longtime2}), at long times
\begin{equation}\label{thermalphi}
 \overline{{\rm Tr}(\hat{\rho}(t)\hat{\Phi})}
 \longrightarrow \int_{- \pi}^\pi\frac{{\rm d}\phi}{2 \pi}\frac{\Phi(\phi)}{({\rm e}^{-\beta t}\cos^2 \phi+{\rm e}^{\beta t}\sin^2 \phi)}\,.
\end{equation}
Angle $\phi$ is defined with respect to the canonical pair $\hat{p} \leftrightarrow p$ and $\hat{q}\leftrightarrow q$ via the variables $(x,y)$.  But the physical variables are $(X,y)$, equation (\ref{coordns}). In particular we may define the physical angle as $\overline{\phi}$ such that
\begin{equation}\label{tan}
  \tan\overline{\phi} = {\rm e}^{\beta t}\tan{\phi.}
\end{equation}
Thus $\overline{\phi}$ is a function of $\phi$, with parametric dependence on $\beta t$. Differentiating both sides of (\ref{tan}) with respect to $\overline{\phi}$ and rearranging terms gives
\begin{equation}\label{angdiff}
  \frac{{\rm d}\phi}{{\rm d}\overline{\phi}}= \big({\rm e}^{-\beta t}\cos^2 \phi+{\rm e}^{\beta t}\sin^2 \phi\big)=
  \frac{1}{\big({\rm e}^{-\beta t}\sin^2 \overline{ \phi}+{\rm e}^{\beta t}\cos^2\overline{\phi}\big)}\,\,,
\end{equation}
so that we can also make the association $\hat{\Phi}\longleftrightarrow \Phi(\overline{\phi})$,
and write
\begin{equation}\label{phibarave}
  \overline{{\rm Tr}( \hat{\rho}(t) \hat{\Phi})}\longrightarrow
  \int_{- \pi}^\pi\frac{{\rm d}\overline{\phi}}{2 \pi}\,\, \Phi(\overline{\phi})\,,
\end{equation}
corresponding to a random distribution in physical angle $\overline{\phi}$. On the other hand from the form of (\ref{thermalphi}) at long times the distribution of $\phi$ becomes strongly concentrated near $\phi=0$.

From (\ref{thermalphi}), (\ref{tan}), and (\ref{angdiff}) it is clear that at long times, whatever the initial state $\hat{\rho}(0)$ may be, the expectations of $\hat{\phi}$ and of $\hat{\overline{\phi}}$ vanish.  Figure 2 shows for three illustrative cases, computed using the full propagator (\ref{osc}) in (\ref{aveop}), the approach of the expectation of $\hat{\phi}$ to zero as functions of time when the initial state is the ground state $|h_0\rangle\langle h_0|$, equation(\ref{ground}). They are: curve A \{$D=\frac{2 \theta}{\hbar\omega}=1000$ and $B = \frac{\beta}{\omega}=0.02$\}; curve B \{$D=10$ and $B=0.1$\}; curve C \{$D=5$ and $B=0.05$\}. Curve A especially corresponds to high temperature and low damping such that their product is 20.  This is nearly identical to the case $\beta=0$ discussed in reference \cite{smith15}.\footnote{The value of parameter $N_o\equiv\mu/(m \omega^2 \hbar)$ in the Figure of that paper should have been stated as $N_o=20$.}

More generally, in the Weyl correspondence, for any angle operator $\hat{\Phi} \longleftrightarrow \Phi(\phi)$, equations (\ref{weyl1}) and (\ref{weyl}) give $\hat{\Phi}$.  For any such function of angle only it can be shown \cite{dhsbook, dhsrims1, dhs95} that its matrix elements with respect to harmonic oscillator eigenstates \{$|h_n\rangle, n=0,1,2\ldots$\} are
\begin{equation}\label{phimn2}
    \langle h_m|\hat{\Phi}| h_n\rangle =
{\rm i}^{m-n}\,g_{m,n}
\int_{- \pi}^\pi \frac{{\rm d}\phi}{2 \pi}\,\Phi(\phi)\, {\rm e}^{{\rm i}(n-m)\phi}\,.
\end{equation}
Here $g_{m,n}$ is the real symmetric matrix
\begin{equation}\label{gmn1}
    g_{m,n} = 2^{-\frac{|m-n|}{2}}\,
    \frac{\Gamma \left( \frac{n_\ell}{2}+ s_\ell\right)}
    {\Gamma \left( \frac{n_g}{2}+ s_\ell\right)}\sqrt{\frac{n_g !}{n_\ell
    !}}
\end{equation}
with
\begin{equation}\label{gmn2}
   s_{\ell} = \left\{ \begin{array}
   {r@{\quad n_\ell}l}
   1/2& \,\,\, \mbox{even}\\
    1\,\,\, & \,\,\,\mbox{odd}
    \end{array}\right.
\end{equation}
and $n_\ell \,(n_g)$ is the lessor (greater) of the pair $(m,n)$.

We may also ask for matrix elements of the operator corresponding to $\overline{\phi}$ which, by equation (\ref{tan}), is a function of $\phi$ only, with $\beta t$ as a parameter. Then
\begin{equation}\label{phihat1}
  \langle h_m|\hat{\overline{\phi}}|h_n\rangle = {\rm i}^{m-n}\,g_{m,n}
\int_{- \pi}^\pi \frac{{\rm d}\phi}{2 \pi}\,\overline{\phi}(\phi)\, {\rm e}^{{\rm i}(n-m)\phi}\,.
\end{equation}
When this is integrated by parts, use made of the fact that $\overline{\phi}(\pm \pi)= \pm \pi$, and recourse made to equation (\ref{angdiff}) one finds upon rearranging that, when $n=m$, $\langle h_m|\hat{\overline{\phi}}|h_n\rangle$ vanishes and when $n\neq m$,
\begin{equation}\label{phihat2}
  \langle h_m|\hat{\overline{\phi}}|h_n\rangle = {\rm i}^{n-m-1} \frac{g_{m,n}}{{(n-m)}}
  \Big\{1-\int_{- \pi}^\pi \frac{{\rm d}\phi}{2 \pi} \frac{{\rm e}^{{\rm i}(n-m)\phi}}{\big({\rm e}^{-\beta t}\cos^2 \phi+{\rm e}^{\beta t}\sin^2 \phi\big)}\Big\}\,.
\end{equation}
The imaginary part of the second term in curly brackets vanishes by symmetry. What remains can be re-expressed as an integral over the range $(0,\pi/2)$ and evaluated by using tables, eg \cite{gradsht}.  The result is
\begin{equation}\label{phihat3}
  \langle h_m|\hat{\overline{\phi}}|h_n\rangle = (1-\delta_{m,n})\, {\rm i}^{n-m-1} \frac{\overline{g}_{m,n}}{{(n-m)}}
\end{equation}
where
\begin{equation}\label{phihat4}
  \overline{g}_{m,n} = g_{m,n}
  \Big\{1-\sigma_{|n-m|}\, \big(\tanh(\beta t/2)\big)^{|n-m|/2}\Big\}\,,
\end{equation}
with
\begin{equation}\label{phihat5}
  \sigma_{n} = \left\{ \begin{array}
   {r@{\quad n}l}
   1& \,\,\, \mbox{even}\\
    0 & \,\,\,\mbox{odd}\,.
    \end{array}\right.
\end{equation}
One might be forgiven for calling calling $\hat{\phi}$ the {\it canonical} phase operator because it is the Weyl quantization of angle in the phase plane defined (effectively) by the canonical coordinates $(p,q)$.  It is a bounded self-adjoint operator discussed at length in \cite{dhsbook} and \cite{dhsrims1}, but properties of the operator $\hat{\overline{\phi}}$, which is based on the physical parameters $(P,q)$, where $P=m\dot{q}=p\, {\rm e}^{-\beta t}$, are open questions. We can, however, get some suggestions by computation. In particular, Figure 3 plots the computed eigenvalues of $\hat{\overline{\phi}}$ for several values of $\beta t$ when its matrix, equation (\ref{phihat3}), is truncated to size $150 \times 150$. Note that the canonical operator corresponds to the case $\beta t =0$. The eigenvalues for that case appear to be spread evenly {\it pari passu}, as befits the canonical phase, from $-\pi$ to $\pi$, but that as $\beta t$ increases the spread of values polarises equally between the values $\pm \pi/2$ although, as $\beta t$ increases, the values $\pm \pi$ are limit points.
\subsection{Angle variance}\label{anglevar}
The variance of the canonical quantised angle is
\begin{equation}\label{sigmaone}
  \langle h_m|\hat{\phi}^2 | h_m\rangle =
  \sum_{n=0}^\infty |\langle h_m|\hat{\phi}|h_n \rangle|^2 =
  \sum_{n=1}^m \frac{1}{n^2}\, (g_{m,m-n})^2 +
  \sum_{n=1}^\infty \frac{1}{n^2}\,(g_{m+n,m})^2\,,
\end{equation}
In the correspondence limit, for which $n$ and $m$ are large compared to their difference, $g_{m,n}$ approaches unity. So for large $m$
\[\langle h_m|\hat{\phi}^2 | h_m\rangle \longrightarrow 2 \sum_{n=0}^\infty \frac{1}{n^2}=\frac{\pi^2}{3}\qquad({\rm as}\,\,m\rightarrow\infty), \]
consistent with a random distribution of angle.

The analysis for $\hat{\overline{\phi}}^2$ follows directly.  From (\ref{phihat3}) one has

\begin{equation}\label{sigmatwo}
   \langle h_m|\hat{\overline{\phi}}^2 | h_m\rangle =
  \sum_{n=0}^\infty |\langle h_m|\hat{\overline{\phi}}|h_n \rangle|^2 =
  \sum_{n=1}^m \frac{1}{n^2}\, (\overline{g}_{m,m-n})^2 +
  \sum_{n=1}^\infty \frac{1}{n^2}\,(\overline{g}_{m+n,m})^2\,.
\end{equation}

This expectation is time-dependent, equation (\ref{phihat4}).  When $\beta t$ vanishes it becomes identical to (\ref{sigmaone}), but as $\beta t \rightarrow \infty$ we have the situation that $\overline{g}_{m,n}$ vanishes when $|m-n|$ is even but equals $g_{m,n}$ otherwise. In that case the result is that as $m$ increases,

\[\langle h_m|\hat{\overline{\phi}}^2 | h_m\rangle \longrightarrow\ 2\sum_{n=0}^\infty \frac{1}{(2n+1)^2}=\frac{\pi^2}{4}\qquad({\rm as}\,\,m\,{\rm and}\,\,\beta t\rightarrow\infty).\]
This makes sense for a distribution of angle between the values between the limiting values $\pm\pi/2$---see Figure 3---with equal probability.

We may ask for the expectation of the square of the canonical phase operator, $\hat{\phi}^2$, in the limit of thermalisation, equation (\ref{longtime2}), when damping is weak, $\beta \rightarrow 0$.  In particular,
\begin{equation}\label{phiRphi}
  \overline{{\rm Tr}\big( \hat{\rho}(t) \hat{\phi}^2 \big)}\longrightarrow
  2\int_0^\infty {\rm d}x\, x\, {\rm e}^{-x^2}
  \int_{-\pi}^{\pi}\frac{{\rm d}\phi}{2 \pi}(\hat{\phi}^2)\Big(x \sqrt{\frac{2 \Theta}{\hbar \omega},\phi}\Big)
  \quad\quad({\rm thermal\,\, limit\,\, but }\,\,\beta\rightarrow 0)\,.
\end{equation}
Now by definition, equation (\ref{weyl1}),
\[(\hat{\phi}^2)(R,\phi)= \int_{- \pi}^\pi \frac{{\rm d}\phi}{2 \pi}{\rm Tr}\big(\hat{\Delta}(R,\phi ) \hat{\phi}^2\big)
=\int_{-\pi}^\pi \frac{{\rm d}\phi}{2 \pi} \sum_{m,n=0}^\infty \langle h_m|\hat{\Delta(R,\phi)}|h_n \rangle
\langle h_n|\hat{\phi}^2|h_m\rangle \]
where (\cite{dhsrims1})
\begin{equation}\label{deltamn}
    \langle h_m|\,\hat{\Delta}(R,\phi)\,| h_n\rangle =
    2 (-1)^n \,{\rm i}^{|m-n|}\, 2^{\frac{|m-n|}{2}}\sqrt{\frac{n_\ell !}{n_g
    !}}\,\,{\rm e}^{{\rm i}(n-m)\phi}R^{|m-n|}\, {\rm
    e}^{-R^2}L^{|m-n|}_{n_\ell}(2 R^2)\,.
\end{equation}
Only the diagonal terms $n=m$ survive to give
\[\overline{{\rm Tr}\big( \hat{\rho}(t) \hat{\phi}^2 \big)}\longrightarrow
4\int_0^\infty {\rm d}x\, x\, {\rm e}^{-x^2}\sum_{m=0}^\infty (-)^m \,{\rm e}^{-\frac{2\Theta}{\hbar\omega}x^2}
L_m\left(-\frac{4\Theta}{\hbar\omega}x^2\right)\langle h_m|\hat{\phi}^2|h_m\rangle\,. \]
The integral is standard (\cite{gradsht}).  The result is, where $D=\frac{2 \Theta}{\hbar\omega}$,
\begin{equation}\label{phi2Lim}
  \overline{{\rm Tr}\big( \hat{\rho}(t) \hat{\phi}^2 \big)}\longrightarrow
  \frac{2}{D+1}\sum_{m=0}^\infty \left( \frac{D-1}{D+1} \right)^m \langle h_m|\hat{\phi}^2|h_m\rangle
  \quad({\rm thermal\,\, limit\,\, but }\,\,\beta\rightarrow 0)\,.
\end{equation}
Consider the high temperature limit of this result, for which $(D-1/(D+1)$ approaches unity, The sum then becomes dominated the increasing number of terms for which $\langle h_m|\hat{\phi}^2|h_m\rangle \rightarrow \pi^2/3$. So that, as $D$ increases we have
\[\lim_{D\rightarrow\infty}\overline{{\rm Tr}\big( \hat{\rho}(t) \hat{\phi}^2 \big)}=\frac{\pi^2}{3}\,. \]
This is characteristic of a random distribution of phase. It is consistent with the analysis of \cite{smith15} which had the white noise driving force $\mu$ but no friction $\beta$.  However, for thermalization to occur in this analysis, equation (\ref{thermalxy}), we require $\mu = 2 m \beta \Theta$, equation (\ref{mu}), so to retain $\mu$ at all we must keep the product $\beta \Theta$  finite.
\section{Oscillator energy}\label{R}
Consider the operator function
\[\hat{O} =  {\rm exp}(-B \hat{E}_{\rm osc})\,, \]
where $\hat{E}_{\rm osc}$ is the oscillator's physical energy operator (noting equation (\ref{phasevars}))
\begin{equation}\label{physenery}
  \hat{E}_{\rm osc}= \frac{\hat{P}^2}{2m}+\frac{m \omega^2}{2}\hat{q}^2
 = {\rm e}^{-2 \beta t} \frac{\hat{p}^2}{2m}+\frac{m \omega^2}{2}\hat{q}^2
\end{equation}
Then, for purposes of operator algebra and the Weyl transform we can scale $m$ and $\omega$ such that $\overline{m}\equiv m \,{\rm e}^{2 \beta t}$ and $\overline{\omega}\equiv \omega  \, {\rm e}^{- \beta t}$ so that
\[\hat{E}_{\rm osc}= \frac{\hat{p}^2}{2\,\overline{m}}+\frac{\overline{m}\,\overline{ \omega}^2}{2}\,\hat{q}^2\,. \]
Correspondingly we can define
\[(\overline{x},\overline{y})\equiv (\frac{p}{\hbar \overline{\alpha}},\overline{\alpha}q) \]
where $\overline{\alpha}^2 \equiv \overline{m}\,\overline{\omega}/\hbar = \alpha^2 \,{\rm e}^{\beta t}$. Now for the basic harmonic oscillator Hamiltonian $\hat{H}=\frac{\hat{p}^2}{2m}+\frac{m \omega^2}{2}\hat{q}^2$ the Weyl correspondence for ${\rm exp}(-B \hat{H})$ is (\cite{smith15})
\begin{equation}\label{hosc}
  {\rm exp}(-B \hat{H}) { \longleftrightarrow}
 \frac{1}{\cosh \left(\frac{\hbar \omega B}{2}\right)}
{\exp}\left[-R^2 \tanh\left(\frac{\hbar \omega B}{2}\right)\right]\,.
\end{equation}
Thus, defining $\overline{R}^2\equiv\overline{x}^2+\overline{y}^2= R^2({\rm e}^{- \beta t}\cos^2 \phi+{\rm e}^{\beta t}\sin^2\phi)$, we can write an expression for the Weyl transform of ${\rm exp}(-B \hat{E}_{\rm osc})$ by replacing in (\ref{hosc}) $R$ by $\overline{R}$ and $\omega$ by $\overline{\omega}$.
The integrals in (\ref{longtime2}) are straightforward, recognizing $\frac{{\rm d}\overline{\phi}}{{\rm d}\phi}$, equation (\ref{angdiff}), gives
\[\overline{{\rm Tr}\big( \hat{\rho}(t){\rm exp}(-B \hat{E}_{\rm osc})  \big)}\longrightarrow
\frac{1}{\cosh(\frac{\hbar\omega}{2}B {\rm e}^{-\beta t})+\frac{2 \Theta}{\hbar \omega}{\rm e}^{\beta t}\tanh(\frac{\hbar\omega}{2}B {\rm e}^{-\beta t})}\,. \]
Finally, as $\beta t \longrightarrow \infty$ this becomes
\[[\overline{{\rm Tr}\big( \hat{\rho}(t){\rm exp}(-B \hat{E}_{\rm osc})  \big)}\longrightarrow
\frac{1}{1+B \Theta}= \frac{1}{1+B kT}\,.\] This is the the classical result for a thermalized harmonic oscillator.
\section{Discussion}\label{disc}
By adding the friction force, this paper generalizes the discussion in \cite{smith15} of an harmonic oscillator acted upon by solely by an external white noise force. In transcribing the classical Brownian dynamics to a semi-classical quantum mechanics it has proved efficient to use the Wigner/Weyl formalism.  The result is semi-classical of course because the external forces are represented by c-numbers, with the twist that the white noise force is stochastic. Though phenomenological the theory can be used to model quantum effects of an oscillator interacting with a heat bath.

Inclusion of friction in the model means that although the Hamiltonian still generates time translation it is not the physical energy.  At long times, with friction acting, the oscillator forgets its initial state and thermalizes, (\ref{thermalxy}) with (\ref{mu}). Section \ref{transprob} considers the time-dependence of the probability that the oscillator remains in its ground state.  The effect of friction is to drive this probability to zero exponentially. Figure 1 shows this.

Section \ref{phase} considers implications for the phase operator $\hat{\phi} \longleftrightarrow \phi$, defined by equations (\ref{phasedef1}) and (\ref{phi1}), and more generally for operators $\hat{\Phi} \longleftrightarrow \Phi(\phi)$. At long times ($\beta t$ large) the distribution of $\phi$ peaks strongly towards zero, equation (\ref{thermalphi}). But $\hat{\phi}$ is defined from the canonical variables $(\hat{p},\hat{q})\longleftrightarrow(p,q)$.  Instead one might define a physical angle operator $\hat{\overline{\phi}}$ in terms of $(m \dot{q},q)\longleftrightarrow ({\rm e}^{- \beta t}\hat{p},\hat{q})$ whose distribution at long times becomes random.  Figure 2 shows the approach to zero in time of the expectation of $\hat{\phi}$ when the oscillator starts in its ground state.

Angle operators $\hat{\phi}$ and $\hat{\overline{\phi}}$ can be represented by the set of their matrix elements between the standard harmonic oscillator states $h_m$, where $m=0,1,2,\cdots$. Figure 3 shows the spread of their approximate eigenvalues when these matrices are truncated to sizes $150 \times 150$. A curious feature of $\hat{\overline{\phi}}$  is that its spectrum depends on time. Consistent with the results of \cite{smith15}, the variance ${\rm Tr}\big( \hat{\rho}(t) \hat{\phi}^2 \big)$ approaches $\pi^2/3$ (as for a random distribution of phase) in the limits $\beta\rightarrow 0$ and $kT \rightarrow \infty$ such that $\mu \equiv 2 m \beta k T$ remains finite.

Section \ref{R} considers for long times the expectation of the generating function $\hat{O} \equiv  {\rm exp}(-B \hat{E}_{\rm osc})$ where $B$ is a c-number parameter and $\hat{E}_{\rm osc}$ is the oscillator's physical energy, equation (\ref{physenery}).  As $\beta t\rightarrow \infty$ this becomes identical with the classical result.

\newpage
\section{Figures}
\begin{figure}[h]
\begin{center}
\includegraphics{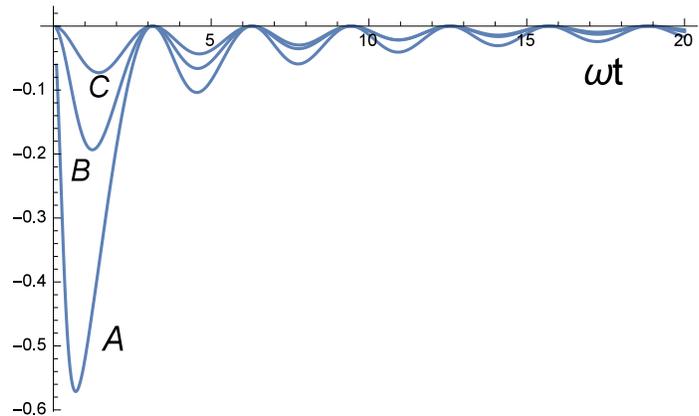}
\caption{\label{PhaseAve} The expectation of the Weyl quantized phase, equation (\ref{phi1}), when the oscillator is initially in its ground state. For curve ``A", $D=\frac{2 \theta}{\hbar\omega}=1000$ and $B=\frac{\beta}{\omega}=0.02.$  For curve ``B", $D=10$ and $B=0.05$.  For curve ``C'' $D=5$ and $B=0.05$.
Curve ``A" very nearly reproduces Figure 1 of reference \cite{smith15} for which $\beta$ is strictly zero and $N_o \equiv \mu/(m \omega^2 \hbar)=20$.}
\end{center}
\end{figure}
\begin{figure}[ht]
\begin{center}
\includegraphics{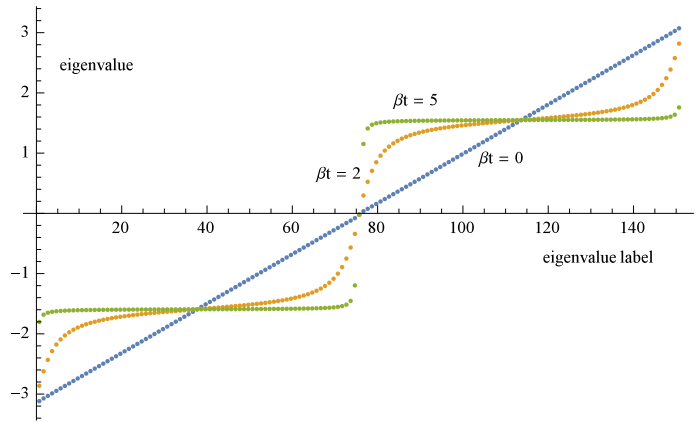}
\caption{Numerical computation of spectrum of $\hat{\overline{\phi}}$, equation (\ref{phihat2}), as approximated by a $150 \times 150$ matrix, equation (\ref{phihat3}), for $\beta t = (0,2,5)$.  The case $\beta t = 0$ reproduces the spectrum for the canonical Weyl quantized phase operator $\hat{\phi}$. }
\end{center}
\end{figure}
\end{document}